\documentclass{vgtc}                          % final (conference style)
\ifpdf%                                % if we use pdflatex
  \pdfoutput=1\relax                   % create PDFs from pdfLaTeX
  \usepackage{graphicx}                % allow us to embed graphics files
  \DeclareGraphicsExtensions{.pdf,.png,.jpg,.jpeg} % for pdflatex we expect .pdf, .png, or .jpg files
\else%                                 % else we use pure latex
  \usepackage{graphicx}                % allow us to embed graphics files
  \DeclareGraphicsExtensions{.eps}     % for pure latex we expect eps files
\fi%

\graphicspath{{figures/}{pictures/}{images/}{./}} % where to search for the images
\usepackage[usenames,dvipsnames]{color}
\usepackage{microtype}                 % use micro-typography (slightly more compact, better to read)
\PassOptionsToPackage{warn}{textcomp}  % to address font issues with \textrightarrow
\usepackage{textcomp}                  % use better special symbols
\usepackage{mathptmx}                  % use matching math font
\usepackage{times}                     % we use Times as the main font
\usepackage{cite}                      % needed to automatically sort the references
\usepackage{tabu}                      % only used for the table example
\usepackage{booktabs}                  % only used for the table example
\onlineid{0}

\vgtccategory{Research}

\vgtcinsertpkg

\title{Judgment as Coordination: A Joint Systems View of Visualization Design Practice}

\author{Paul C. Parsons\thanks{e-mail: parsonsp@purdue.edu}\\ %
        \scriptsize Purdue University %
\and Arran Ridley\thanks{e-mail: arran.ridley@monash.edu}\\ %
     \scriptsize Monash University, Indonesia % \\ %
     }

\abstract{
Professional visualization design has become an increasingly important area of inquiry, yet much of the field’s discourse remains anchored in researcher-centered contexts. Studies of design practice often focus on individual designers’ decisions and reflections, offering limited insight into the collaborative and systemic dimensions of professional work. In this paper, we propose a systems-level reframing of design judgment grounded in the coordination and adaptation that sustain progress amid uncertainty, constraint, and misalignment. Drawing on sustained engagement across multiple empirical studies—including ethnographic observation of design teams and qualitative studies of individual practitioners—we identify recurring episodes in which coherence was preserved not by selecting an optimal option, but by repairing alignment, adjusting plans, and reframing goals. We interpret these dynamics through the lens of Joint Cognitive Systems, which provide tools for analyzing how judgment emerges as a distributed capacity within sociotechnical activity. This perspective surfaces often-invisible work in visualization design and offers researchers a new conceptual vocabulary for studying how design activity is sustained in practice.
}

\CCScatlist{
  \CCScat{Human-centered computing}{Visu\-al\-iza\-tion}{}{}
}

\begin{document}



\maketitle

\section{Introduction}
The visualization research community has shown increasing interest in professional design practice—not just as a site for deploying research innovations, but as a domain worthy of study in its own right. Researchers have examined how visualization work unfolds in real-world contexts, where collaboration, ambiguity, and evolving constraints shape design trajectories~\cite{parsons_understanding_2021, lee-robbins_client-designer_2024, walchshofer_transitioning_2024, stokes_its_2025, bigelow_reflections_2014, alspaugh_futzing_2019, baigelenov_how_2025}. While this shift has yielded several rich accounts of practice, many of the conceptual tools used to describe visualization design—process models, decision frameworks, evaluation protocols—continue to reflect the goals and conditions of research-centered work more than those of professional practice.

This paper contributes to a growing effort to expand the field’s conceptual vocabulary for studying professional design. Prior work has shown that judgment plays a central role in shaping how visualization practitioners frame problems, navigate trade-offs, and interpret constraints~\cite{parsons_design_2020}. Across multiple empirical studies—including an ethnographic investigation of team-based visualization projects in a design studio, and complementary studies of individual practitioners—we observed that design judgment rarely operates as an isolated cognitive act. Instead, it emerges through distributed coordination: across people, tools, and organizational systems. These observations motivate a systems-level reframing of design judgment as an adaptive capacity enacted through joint activity.

We develop this perspective using concepts from Joint Cognitive Systems (JCS) theory~\cite{woods_joint_2006,hollnagel_joint_2005}, which offers analytic tools for understanding how work is sustained under complexity, uncertainty, and constraint. Rather than focus on isolated decision points or process steps, we draw attention to episodes in which progress in the design process depends on maintaining shared understanding, repairing misalignment, or reframing goals. Our aim is not to report new empirical findings, but to reflect on insights from prior work to foreground a dimension of design practice that is often invisible: how judgment operates as a sociotechnical system’s capacity to preserve coherence and adapt over time.

\section{Background}

\subsection{Judgment in Design Practice}

Design judgment refers to the capacity to navigate complexity, ambiguity, and competing values in the pursuit of purposeful change. Judgment is central to design because it enables practitioners to act meaningfully in situations that cannot be guided by formal models~\cite{nelson_design_2012}. Judgment is not simply about choosing between predefined options—it is about framing problems, appreciating context, interpreting values, and making situated decisions under constraint. As Dunne~\cite{dunne_professional_1999} argues, professional judgment involves mediating between general principles and the particular demands of a situation, requiring flexibility, reflection, and adaptation—qualities that formalized models often struggle to capture. %Our systems-level framing builds on this insight by examining how such judgment is enacted collaboratively within sociotechnical practice.

This view of judgment has been extended into applied design contexts, including instructional design~\cite{lachheb_role_2021} and data visualization~\cite{parsons_design_2020}. Design judgments of visualization practitioners have been characterized as layered and context-sensitive, encompassing framing, appreciative, and instrumental types, among others~\cite{parsons_design_2020} . Our work builds on this foundation by shifting attention from the types of judgment to the dynamics through which judgment is enacted collaboratively in practice. We treat judgment not as an internal faculty of the designer, but as a distributed and adaptive capacity that emerges through joint activity.

\subsection{Professional Visualization Design Practice}
A growing body of research has highlighted the complexity of visualization design practice, showing that designers must continuously align with shifting stakeholder goals, negotiate constraints, adapt to organizational structures, and engage in reflection-in-action~\cite{lee-robbins_client-designer_2024, walchshofer_transitioning_2024, walny_data_2020, meyer_reflection_2018,dhawka_social_2025, parsons_understanding_2021}. Together, this work foregrounds the situated and improvisational activity that characterizes design work. Similar observations have been made in other design domains, where collaboration depends critically on organizational structures, communication flows, and the continual repair of shared understanding~\cite{chiu_organizational_2002}. Researchers have highlighted visualization design judgment as a layered, context-sensitive, and shaped by reflection, improvisation, and coordination~\cite{parsons_design_2020, alspaugh_futzing_2019, bigelow_reflections_2014}. Studies of dynamic projects (e.g., designing pandemic dashboards) have illustrated how coordination is essential for sustaining momentum under pressure and uncertainty~\cite{zhang_visualization_2022}. Others have examined how mutual understanding evolves in client-designer collaboration, or how design artifacts serve to stabilize framing across actors and time~\cite{lee-robbins_client-designer_2024, stokes_its_2025}.

This prior research has deepened our understanding of practice, but few studies offer theoretical tools for analyzing the distributed dynamics that sustain progress amid misalignment, shifting constraints, and organizational complexity. Our work contributes to this trajectory by introducing a systems-theoretic framing of design judgment, grounded in JCS theory. We extend existing empirical work by offering a vocabulary for interpreting coordination breakdowns, adaptive trade-offs, and reframing episodes as core mechanisms of professional visualization work.

\subsection{Coordination and Adaptation in Practice}

Joint Cognitive Systems (JCS) theory~\cite{woods_joint_2006, hollnagel_joint_2005} provides a lens for understanding how complex work is done through coordination across people, tools, and representations. Rather than foregrounding individual decisions, JCS emphasizes how systems sustain performance by maintaining common ground, adapting to evolving conditions, and repairing breakdowns. Resilience Engineering (RE)~\cite{hollnagel_resilience_2006} extends this view, introducing concepts like \textit{adaptive capacity}, \textit{sacrifice judgments}, and \textit{graceful extensibility} to describe how systems remain functional amid uncertainty, constraint, and change.

Other cognition frameworks such as Distributed Cognition (DCog), Activity Theory (AT), and Actor-Network Theory (ANT) offer tools for analyzing sociotechnical systems. While DCog emphasizes distributed processes, AT highlights mediated, goal-directed activity, and ANT traces associations among human and non-human actors, JCS and RE focus more directly on adaptation under constraint---how systems sustain coherence and extensibility amid breakdowns and evolving demands. This emphasis aligns with the realities of visualization practice, where progress often hinges on practitioners' ability to reorient, adapt, and preserve momentum.

Although JCS and RE were originally developed in safety-critical fields, they are increasingly being adopted in domains like software infrastructure~\cite{robbins_resilience_2012}. Recent reflections by RE researchers~\cite{herrera_resilience_2024} advocate for applying these ideas more broadly to sociotechnical systems where coordination and adaptation are central. We argue that professional visualization design is one such domain: practitioners must make progress despite ambiguous goals, shifting requirements, limited resources, and imperfect communication across roles and tools. In this context, the JCS/RE perspective offers analytic leverage for understanding design work that does not proceed linearly or transparently. It helps explain how teams respond to misalignment, navigate trade-offs, and sustain momentum when priorities shift or breakdowns occur. By focusing on coordination and adaptation, this framing makes visible forms of judgment that are often taken for granted or omitted from formal design models. To date, these frameworks have not been applied to research on visualization design. We propose that they offer a valuable foundation for analyzing how judgment emerges not as individual insight, but as a system-level capacity to maintain coherence and adapt under constraint.

\section{Empirical Basis and Interpretive Approach}

Our insights are grounded in sustained empirical engagement with professional visualization design practice, spanning two complementary research programs. One author conducted an ethnographic study within an award-winning European design studio, observing team-based visualization work across multiple client projects~\cite{ridley_evaluating_2020,lee-robbins_client-designer_2024,ridley_sociocultural_2021}. At the time (2018-2019), the studio employed 28 people and produced a range of visualization design work, from large-scale dashboards for multinational corporations to style guides for charities. The other author has led a series of studies involving 45 data visualization practitioners, using methods such as design challenges, diary studies, and interviews to examine how designers navigate framing, ambiguity, and judgment in their practice~\cite{parsons_understanding_2021,parsons_fixation_2021,parsons_design_2020,baigelenov_how_2025,parsons_data_2020}. Participants were primarily based in North America and Europe, and worked across a wide range of professional contexts, including freelance roles in journalism, public policy, health, education, and the non-profit sector, as well as in-house positions within large organizations and data consultancy firms. Rather than report the results of a single empirical study, this paper synthesizes recurring insights across these lines of work to develop a conceptual reframing of design judgment. To guide our analysis, we draw on concepts from JCS/RE, which offer tools for interpreting coordination, adaptation, and the maintenance of shared understanding under real-world conditions.

While the analytic emphasis of this paper is grounded in team-based visualization work—particularly as documented in the ethnographic study—both research programs surfaced recurring patterns of joint activity. Our broader engagement with individual practitioners also informed our theoretical framing. Although the focus is on collaboration, we observed that JCS/RE dynamics, such as coordination repair and adaptive trade-offs, also emerged in individual work situated within sociotechnical systems. Across both contexts, we repeatedly encountered situations in which design progress hinged not on producing the “right” solution, but on navigating misalignment, absorbing uncertainty, and maintaining forward momentum. This interpretive synthesis follows calls within the visualization community for reflexive and theoretically engaged accounts of practice~\cite{meyer_reflection_2018, meyer_criteria_2020}. Our goal is not to generalize statistically or document exhaustive cases, but to surface how judgment operates through coordination, adaptation, and reframing.

\section{Patterns of Judgment as Joint Activity}

We present three interrelated patterns that illustrate how design judgment in visualization emerges through joint activity, synthesized from our empirical engagement with professional design practice, particularly grounded in team-based observations. Guided by concepts from JCS/RE, they highlight how coordination, adaptation, and reframing enable progress amid ambiguity and constraint.

\subsection{Coordination Breakdown and Repair}
\label{sec:coordinationbreakdown}
From a JCS perspective, \textit{coordination} refers to the management of interdependent actions within a system (e.g., across people and tools), while \textit{common ground} refers to the shared understanding that enables such coordination. When common ground breaks down—due to misaligned assumptions or unclear goals—coordination falters, and judgment must be enacted to restore coherence and progress. In one project, conflicting feedback from stakeholders left the design team uncertain about how to proceed. Rather than escalate the inconsistency, the team convened a realignment meeting to reconstruct a workable frame of reference. This is an instance of common ground maintenance, where coordination is repaired through interpretive alignment. Later that week, a senior designer drafted a clarification message and shared it with teammates. The team’s internal chat log shows careful discussion about wording, tone, and the risk of appearing confrontational. The final message tested whether feedback had shifted, while preserving ambiguity to reduce risk. This exchange reflects how representational coordination can help restore common ground and re-enable productive collaboration.

An extended breakdown occurred in a meeting between client-side infrastructure leads and the visualization team. A key question—\textit{``Who is responsible for what?"}—arose repeatedly, particularly regarding the API layer and the data model underlying the dashboards. Different teams believed they owned different layers of the stack, but no shared artifact or formal clarification existed. The ambiguity about handoffs between data, middleware, and presentation layers stalled decision-making for several weeks. This episode reflects how coordination failures are often not just interpersonal but structural. The lack of mutual visibility into roles and timing requirements created a mismatch in expectations. From a JCS perspective, this is a failure to maintain common ground, exacerbated by organizational silos and diverging timelines. What appears as a technical delay is in fact a breakdown in collective judgment about system design responsibility. In another case, uncertainty about stakeholder needs and division of responsibility persisted well into project planning. Early coordination documents and planning notes reflect unclear expectations around who was accountable for key design decisions and integration tasks. Lacking definitive guidance, the team adopted a strategy of provisional framing---making educated assumptions about priorities and deliverables to keep momentum. This episode highlights the role of joint activity not just in producing artifacts, but in constructing the shared interpretive ground on which judgment becomes actionable.

\subsection{Sacrifice Judgments and Adaptive Trade-offs}

While \textit{sacrifice judgment} in RE has been characterized in relation to safety and risk (i.e., sacrificing efficiency to maintain safety)~\cite{hollnagel_resilience_2006}, the concept can be used more broadly to refer to using judgment to sacrifice one goal in service of another~\cite{shukla_coping_2025}. Such judgments are not failures, but adaptive moves to maintain viability. Designers often make trade-offs about where to invest their limited time and energy. For example, in one case, practitioners documented a decision to remove a planned interaction feature after observing that it confused users during informal testing. While the feature was seen as potentially valuable, it was dropped to preserve resources for more critical needs. Across the projects, such trade-offs reflect a pragmatic orientation toward progress: when uncertainty and time pressure make it difficult to satisfy all goals, practitioners made situated judgments to prioritize momentum over refinement. 

Another example came from internal preparation for a presentation. One designer pushed to refine several secondary visuals, but the team ultimately deprioritized them, anticipating that they would have limited impact on the client conversation. This judgment reduced design time and redirected attention to the summary visual that was expected to dominate the client conversation. Here, sacrifice judgments were shaped not only by time constraints but also by expectations about audience attention and framing. In one project, a mid-fidelity prototype was built using static data rather than the originally intended live feed. Planning documents show that this choice was framed as temporary—a way to avoid integration delays while maintaining design momentum. Though positioned as reversible, the trade-off reflected a broader judgment about the balance between fidelity and progress.

Sacrifice judgments also emerged in relation to managing scope. In one project, a planned comparison across multiple programs was dropped in favor of focusing on a single outcome. Rather than revisit the data pipeline or restructure the visualization, the team adjusted the narrative to align with what could be delivered within current constraints. This was not just a resource trade-off, but a reframing of what the visualization was meant to communicate. Such adaptations exemplify what RE terms \textit{graceful extensibility}: the ability of a system to stretch its goals and resources without losing coherence~\cite{woods_theory_2018}. The team’s ability to let go of features or shift emphasis illustrates how judgment sustains progress—not through ideal solutions, but through adaptive responsiveness to context. These kinds of judgments are often invisible in formal design models or evaluation criteria, yet they shape what gets built, how quality is defined, and which compromises are accepted in professional visualization work.

\subsection{Reframing and Goal Re-alignment}
Perhaps the clearest expression of judgment as joint activity came in episodes of \textit{reframing}. In our context, reframing refers to the process of revising how a design problem, goal, or purpose is understood—often in response to evolving constraints, cues, or stakeholder input. In one project, designers initially understood the client’s goal as surfacing comparative performance metrics, but gradually came to see the primary purpose as supporting a narrative of program value and legitimacy. This shift emerged through repeated client feedback that emphasized positive framing, downplayed uncertainties in the data, and prioritized visual polish over analytical nuance. Retrospective notes document the team’s evolving interpretation of project goals, including concerns about overpromising and a recognition that messaging was taking precedence over strict accuracy. These cues prompted the team to regroup and adjust their design direction, focusing more on credible storytelling and alignment with the client’s strategic goals.

In another project, reframing occurred when an interactive component initially conceived as a filter was reinterpreted as a narrative progression tool. Designers noticed that stakeholders followed a particular order when exploring the data. The team redesigned the interaction to support this narrative flow, changing both its appearance and behavior. A subtler reframing episode appeared when early sketches led clients to focus on interface layout instead of the data itself. The team shifted the review structure to emphasize data model discussion before revisiting the UI.

These reframing moves can involve shifts in what a project is understood to be ``about." They redirect effort, justify discarding prior work, and redefine what success looks like. This is not design judgment as selection from a menu, but as a reinterpretation of what matters. In many instances we witnessed, reframing did not simply reflect a shift in interpretation---it reshaped subsequent decisions, focus, and evaluation criteria. By recognizing subtle cues and realigning project narratives, what counted as success was revised.

\subsection{Examples}
We present a vignette drawn from our observations, followed by a hypothetical scenario designed to illustrate how JCS/RE concepts might manifest in real-world visualization contexts.

\textbf{Vignette:} In the second phase of one project (see Section \ref{sec:coordinationbreakdown}), progress stalled during integration between the dashboard and the client’s data infrastructure. The design team assumed the infrastructure group would manage the API endpoints; the infrastructure team assumed the opposite. Program managers had already committed to deadlines without confirming interdependencies. A joint meeting surfaced incompatible assumptions about roles and timing, creating a breakdown in shared understanding (i.e., a failure to maintain common ground). Judgment could not be exercised until coordination was repaired. In response, the team created a shared coordination document mapping responsibilities and convened a follow-up alignment session (representational coordination). Only then did design work resume. What appeared as a technical delay was, in fact, a systemic coordination breakdown, resolved not by deeper expertise, but by restoring mutual awareness and shared interpretive ground (recovery of judgment capacity through adaptive coordination repair).

\textbf{Hypothetical Scenario:} A team tasked with designing a public air quality dashboard may find themselves collaborating with city officials, data engineers, and environmental justice advocates. While the team might initially frame the dashboard as a tool for temporal data analysis and geographic comparison, stakeholders could express diverging priorities—such as emphasizing accessibility, surfacing disparities, or ensuring visual polish. This kind of misalignment could reflect a breakdown in shared understanding (i.e., failure to maintain common ground). Rather than pausing the project entirely, the team might respond by convening a low-fidelity sketching workshop to surface stakeholder interpretations (representational coordination). Through this process, shared values—like clarity, fairness, and public trust—could emerge to support a reframing of the dashboard’s purpose, shifting it toward a mobile-friendly risk communication tool (goal reframing). To maintain momentum under tight constraints, the team may need to make sacrifice judgments: simplifying interactivity, limiting historical data, or using static samples instead of live feeds. These trade-offs would reflect an effort to preserve essential system functions under evolving demands (i.e., graceful extensibility).

\section{Why Judgment as Joint Activity Matters}

While our perspective is grounded in specific contexts, we believe the core dynamics of coordination repair, adaptive trade-offs, and reframing are relevant across many design settings. We expect this systems-level framing of judgment to be valuable across a range of projects and contexts. In this section, we reflect on the implications of these ideas for visualization research and practice.

\subsection{Extending Practice-Oriented Perspectives}

Recent scholarship in visualization has emphasized that design is shaped by context, collaboration, and iteration~\cite{bigelow_iterating_2017, parsons_understanding_2021, meyer_reflection_2018, alspaugh_futzing_2019, ridley_evaluating_2020}. Our work complements and extends this view by introducing an interpretive lens that focuses on the connective tissue between design actions: how judgment is enacted through joint coordination in response to ambiguity, constraint, and misalignment. We do not argue against established design models, but aim to make visible the informal and often-invisible coordination work that makes structured design processes viable in practice. Treating judgment as a system-level phenomenon broadens the analytical lens, drawing attention to how coherence is maintained and meaning is negotiated across people, tools, and representations. This systems-level view aligns with philosophical accounts of professional practice that emphasize the irreducibility of situated judgment, where action unfolds through improvisation, negotiation, and adaptation rather than following predetermined steps~\cite{dunne_professional_1999}.

\subsection{Implications for Tool and Process Support}

If visualization design hinges on reframing, sacrifice judgments, and coordination repair, tools should do more than support artifact creation—they should scaffold the judgment work that sustains alignment. For reframing, design tools could include lightweight mechanisms to capture evolving rationales. Platforms like Notion or Google Docs are often used informally for this; embedding rationale annotations in design tools could help track shifts in goals or interpretations over time. To support sacrifice judgments, teams might benefit from features that log what was dropped and why. Platforms like Figma already support this informally through comments or hidden layers; dedicated markers for deferred or omitted features could aid memory and alignment. For coordination repair, collaborative environments could surface dependencies or unresolved assumptions. Prompts to clarify roles or ownership—especially at handoff points—could help preempt breakdowns. These features need not formalize every decision, but they can help externalize the adaptive work enabling design processes to remain coherent under constraint.

\subsection{Implications for Research and Pedagogy}

For researchers, this framing invites new forms of analysis that attend to the adaptive, distributed nature of real-world design work. It highlights the importance of studying how teams respond to breakdowns, reinterpret goals, and redistribute attention and responsibility under evolving constraints. These dynamics may be especially salient in multi-stakeholder, longitudinal, or interdisciplinary projects, where priorities shift and communication gaps are common. In addition to reinterpreting practice, this perspective also reshapes how it can be studied. Rather than focusing solely on the choices designers make—such as which chart type was selected or why a particular feature was prioritized—researchers might examine how constraints shaped coordination across roles, or how shared understanding was reestablished following misalignment. Instead of interviewing individuals only about their decision rationale, researchers could trace how teams adapted to infrastructure failures, ambiguous stakeholder signals, or cascading technical dependencies. Rather than analyzing design outcomes in isolation, studies could explore how judgment was scaffolded through representational artifacts, temporal workarounds, or collaborative repair. This approach foregrounds the enabling conditions of design work rather than only its outcomes.

For educators, this work suggests a shift in how decision making in visualization design is conceptualized and taught. Much of the literature treats design as a matter of navigating a predefined ``design space''---i.e., choosing among known options. In contrast, we advocate for teaching judgment as a form of situated coordination, where goals and constraints evolve and must be actively interpreted. Pedagogical approaches might include exercises in reflective trade-off analysis, collaborative framing workshops, pre- and post-mortems, and reflection activities focused on adaptation strategies that sustain project momentum---all of which have not been of focus yet in the literature on data visualization pedagogy. By foregrounding the joint nature of judgment, and embracing a systems-level perspective on visualization design, educators can better prepare students for the complexities of professional practice.

\subsection{Positioning within Broader Conversations}

This work contributes to a growing shift in visualization research toward more reflexive, interpretive, and practice-sensitive approaches to design. Scholars have called for greater attention to the complexities of applied visualization work, including the importance of infrastructural, collaborative, and reflective dimensions often underrepresented in formal models~\cite{meyer_criteria_2020, meyer_reflection_2018,dhawka_social_2025,ridley_evaluating_2020,mendez_bottom-up_2017}. By examining how judgment is enacted through coordination repair and reframing, our work contributes a lens for interpreting reflection-in-action and provides a vocabulary for making visible the tacit work that sustains progress. Other studies have highlighted how visualization teams operate under evolving constraints and shifting priorities~\cite{lee-robbins_client-designer_2024,walny_data_2020}, including in high-pressure or crisis contexts~\cite{zhang_visualization_2022}. In our empirical work, we have observed similar patterns of joint activity and adaptive judgment, suggesting that these dynamics are defining features of professional visualization work. Existing frameworks offer valuable scaffolding for organizing design processes and decisions (e.g., ~\cite{sedlmair_design_2012, mckenna_design_2014,munzner_nested_2009}), yet they often presume stable goals and clearly defined phases. Our work complements these models by surfacing how teams sustain coherence when roles blur, assumptions diverge, or constraints shift unexpectedly. This systems-oriented reframing opens up new theoretical possibilities for the field. By shifting attention from designer decisions to system-level dynamics, JCS/RE encourages researchers to examine how coherence is sustained across distributed agents, tools, and representations. It complements these models by shifting focus from the decisions themselves to the conditions that make decision-making possible. As visualization research increasingly engages with real-world complexity, this framing provides a valuable lens for interpreting the resilience, fragility, and adaptability of design practice. 

\section{Conclusion}

Professional visualization design unfolds through collaborative activity—among designers, clients, tools, and artifacts—under conditions of ambiguity, constraint, and continual change. We have proposed a systems-level framing of design judgment as a capacity enacted through joint activity. Drawing on JCS/RE theory, we examined how teams sustain coherence by repairing coordination, negotiating trade-offs, and reframing goals. While drawn from specific cases, the dynamics we describe are broadly relevant. These insights offer a new lens and vocabulary for studying and teaching visualization design. By focusing on the dynamics of coordination and adaptation, we surface a dimension of design work that is often hidden, yet foundational, to how visualization projects succeed. We hope this perspective supports ongoing efforts to develop methods, models, and tools that better reflect the realities of professional design.

\acknowledgments{
We wish to thank the study participants for their valuable contributions. This work was supported in part by NSF grant \#2146228 and an ESRC White Rose DTP-funded
PhD project (ES/J500215/1).}

\bibliographystyle{abbrv-doi-hyperref}

\bibliography{references}
\end{document}